\documentclass[%
 preprint,
 amsmath,amssymb,
 aps,
]{revtex4-2}
\usepackage[english]{babel} 
\usepackage{graphicx}
\graphicspath{{.},{figures/}}
\usepackage{dcolumn}
\usepackage{bm}

\usepackage{subfigure}

\usepackage{amsmath}
\usepackage{tikz}
\usepackage{tikz-cd}
\usetikzlibrary{positioning,arrows} 
\usetikzlibrary{decorations.pathreplacing} 
\usepackage{tikzsymbols} 

\usepackage{tikz}
\usetikzlibrary{shapes,arrows}
\usepgflibrary{arrows}
\usetikzlibrary{decorations.pathreplacing}
\usetikzlibrary{spy}
\usetikzlibrary{shapes,arrows}
\usetikzlibrary{calc}
\usetikzlibrary{arrows, arrows.meta}
\usepgflibrary{arrows}

\usetikzlibrary{shapes.geometric, arrows}
\tikzstyle{process} = [rectangle, minimum width=1.5cm, minimum height=1cm, text centered, text width=1.5cm, draw=black, thick, fill=black!30]
\tikzstyle{whiteprocess} = [rectangle, minimum width=1.5cm, minimum height=0.5cm, text centered, text width=2.1cm, draw=black, thick, fill=black!0]
\tikzstyle{decision} = [diamond, minimum width=4cm, minimum height=1cm, text centered, text width=3cm, draw=black, fill=black!20]
\tikzstyle{arrow} = [thick,->,>=stealth]

\usepackage{relsize}

\tikzset{fontscale/.style = {font=\relsize{#1}}
    }
    
\usepackage[export]{adjustbox}

\usepackage{scalerel}
\usepackage{stackengine,wasysym}

\begin{document}


\title[A new paradigm for wall-modeled large eddy simulations using the volume-filtering framework]{A new paradigm for wall-modeled large eddy simulations using the volume-filtering framework}

\author{M. Hausmann}
\email{max.hausmann@ovgu.de}
\author{B. van Wachem}
\affiliation{ 
Chair of Mechanical Process Engineering, Otto-von-Guericke-Universit{\"a}t Magdeburg, \\
  Universit{\"a}tsplatz 2, 39106 Magdeburg, Germany
}

\date{\today}

\begin{abstract}
In the present paper, we apply the framework of volume-filtering, initially proposed by \citet{Anderson1967} for particle-laden flows, to large eddy simulations (LES) of wall-bounded flows leading to a new perspective on wall-modeled LES (WMLES) that we refer to as volume-filtered WMLES (VF-WMLES). In contrast to existing wall-models, the VF-WMLES framework does not rely on temporal averaging, does not make a priori assumptions on the pressure gradient, and can be used with a uniform spatial filter, which avoids the appearance of commutation closures in spatial derivatives of the filtered momentum and continuity equation. Volume-filtering is well defined, even close to the wall, and it is shown that a non-zero slip and penetration velocity at the wall is a direct consequence of volume-filtering the flow. With the VF-WMLES concept, new wall models can be directly assessed in a priori and a posteriori studies by comparing the predicted slip and penetration velocities at the wall with the velocities from explicitly volume-filtered direct numerical simulations (DNS). Based on the VF-WMLES concept, we derive an LES modeling strategy that is based on the recently proposed PC-IBM \cite{Hausmann2024b}, a modeling framework based on volume-filtering allowing to couple the flow with arbitrarily shaped solid boundaries using relatively coarse Cartesian fluid meshes. The proposed VF-WMLES is validated with two cases, a turbulent channel flow and a turbulent flow over periodic hills, and shown to accurately predict the mean velocity profiles for both cases.
\end{abstract}

\maketitle

\newpage

\section{Introduction}
Real-world flow problems usually take place in bounded flow domains. Although the concept of periodic boundaries is a useful simplification for some applications, to simulate problems with natural and industrial relevance, the numerical representation of impermeable walls is often required. In direct numerical simulations (DNS), where all relevant length- and time scales of the turbulent flow are resolved, no-slip and no-penetration boundary conditions are typically employed at the walls. Since such a full resolution of turbulent flows requires tremendous computational resources, a DNS is not feasible for most of the applications. A computationally more efficient alternative is to solve only for the presumably more relevant large flow scales, i.e., the spatially low-pass filtered flow scales, and apply an appropriate model for the smaller scales, which is done in large eddy simulations (LES). \\
LES of wall-bounded flows can be categorized in wall-resolved LES, that refine the fluid mesh in the near-wall region to essentially achieve a DNS resolution near the wall (see, e.g., \citet{Frohlich2005}), and the wall-modeled LES (WMLES), that retain a coarse fluid mesh in the whole domain and apply additional modeling near the wall (see, e.g., \citet{Deardorff1970,Schumann1975}). As shown by \citet{Choi2012}, wall-resolved LES can reduce the computational costs significantly compared to a DNS, but in most applications only a WMLES is computationally feasible. To allow a coarse fluid mesh resolution in the near-wall region in WMLES, the velocity is assumed to be filtered with a non-zero filter width, i.e., convoluted with a filter kernel of non-zero support. The problem with retaining a finite filter width near the wall in WMLES is that the convolution operation is not defined in the wall normal direction when the support of the filter kernel exceeds the wall. This leads to wall models that mostly rely on ad hoc interventions to obtain accurate mean statistics. \\
The by far most common approach for WMLES is wall shear stress modeling, where the wall shear stress (typically its statistical or temporal mean) is estimated to formulate a Neumann boundary condition for the streamwise velocity (see, e.g., \citet{Bose2018}). The background is that enforcing the correct wall shear stress in turbulent flows over flat walls guarantees the correct bulk velocity. The mean wall shear stress can be approximated by assuming the mean streamwise velocity profile of the turbulent flow over a flat plate without pressure gradient \cite{Deardorff1970,Schumann1975,Piomelli1989}, or by solving a transport equation for the mean wall-parallel velocity \cite{Balaras1996,Cabot2000,Wang2002b}. Either of the approaches, however, combines mean velocity profiles with the instantaneous filtered velocity present in the LES, two concepts that are incompatible. An even more fundamental issue of wall shear stress models is that 
a boundary condition of the unfiltered velocity, i.e., the unfiltered wall-shear stress, is applied to the filtered velocity field, which can, by definition, not contain such large wave numbers. \\
An alternative to wall shear stress modeling is the dynamic slip wall model, proposed by \citet{Bose2014}, which employs a differential filter with spatially varying filter width. Although the filter width vanishes as the wall is approached, the filtered velocity is not equal to the no-slip and no-penetration boundary condition at the wall but takes a non-zero value. Despite the non-uniform filter width, \citet{Bose2014} and \citet{Bae2019} employ a uniform fluid mesh also near the wall, which may not be fine enough to resolve the large wave numbers that potentially arise in the near-wall region where the filter width is small. Although the consequence of a non-uniform filter on the resulting filtered velocity at the wall is captured correctly with the dynamic slip wall model, it is still assumed that spatial derivative and filtering commute, which is incorrect for non-uniform filters \cite{Ghosal1995,Moser2021,Yalla2021a}. \\
A concept that is related to what is proposed in the present paper is the filtered-wall formulation of \citet{Bhattacharya2008}, where a region with zero velocity is appended outside the flow domain, such that the filtering operation can be carried out even close to the wall. An additional closure term is shown to emerge, which is modeled using a so-called matched buffer approach, enforcing that the filtered velocity at the wall equals the twice filtered velocity. A clear justification or data for this assumption is not provided and no modeling framework is derived that can be practically applied in LES. \\
In the present paper, we apply the concept of volume-filtering, originally proposed by \citet{Anderson1967}, to obtain the equations which govern LES of wall-bounded flows. Volume-filtering redefines the filter as convolution over the domain where fluid quantities are defined. Recent advances in theory \cite{Hausmann2024a} and modeling \cite{Hausmann2024b} in the scope of volume-filtering allow to derive a practically applicable modeling framework for WMLES, that we refer to as volume-filtered WMLES (VF-WMLES). It is shown that a resulting slip and penetration velocity at the wall is a direct consequence of volume-filtering the wall-bounded flow. This allows to assess newly proposed wall models, by comparing the predicted slip and penetration velocity at the wall with the explicitly volume-filtered DNS velocity field. As another consequence of volume-filtering the Navier-Stokes equations (NSE), a momentum source term arises in the volume-filtered momentum equation. The force on the wall resulting from this momentum source equals the integral of the unfiltered wall shear stress and unfiltered pressure over the wall, although the volume-filtered velocity gradient is much smaller than the unfiltered velocity gradient near the wall. The VF-WMLES can be used with a uniform filter, which avoids the necessity of modeling commutation closures that arise in every spatial derivative of the filtered Navier-Stokes equations (FNSE) when employing a non-uniform filter. Furthermore, the VF-WMLES does not make any a priori assumptions on the pressure gradient and is conceptually suitable for curved walls with flow separation. \\
In the present paper, the VF-WMLES is realized with a recently proposed generalization of the immersed boundary method (IBM), the so-called PC-IBM \cite{Hausmann2024b}. The PC-IBM is fundamentally different to previously proposed IBM approaches for WMLES \cite{Roman2009a,Capizzano2011,Chang2014}, which combine classical WMLES strategies with the numerical framework of IBM to allow the simulation of curved walls using a Cartesian fluid mesh. The PC-IBM, which is directly derived from the volume-filtered NSE, allows to consistently represent arbitrarily shaped boundaries on a relatively coarse Cartesian fluid mesh. \\
The remainder of the paper is structured as follows. The concept of volume-filtering is introduced in section \ref{sec:theoryvolumefiltering} and the effect of volume-filtering a wall-bounded flow on the velocity at the wall is demonstrated. Section \ref{sec:modeling} describes how the modeling with the VF-WMLES is realized using the PC-IBM. The VF-WMLES is validated with two cases, a turbulent channel flow and a turbulent flow over period hills, which is provided and discussed in section \ref{sec:results}. Finally, the paper is concluded in section \ref{sec:conclusions}

\section{Theory of volume-filtering }
\label{sec:theoryvolumefiltering}
\subsection{Fundamentals of volume-filtering}
\begin{figure}
    \centering
    \includegraphics[scale=1]{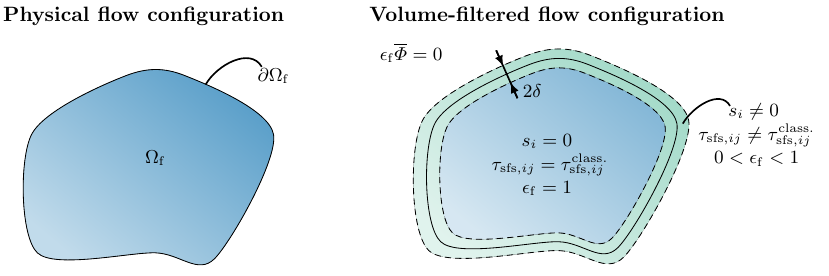}
    \caption{Illustration of volume-filtering the flow field in a confined domain. On the left, the physical flow configuration is shown, which is bounded by a no slip wall, $\partial\Omega_{\mathrm{f}}$. The volume-filtered flow configuration is shown on the right, and extends beyond the boundary $\partial\Omega_{\mathrm{f}}$, but the volume-filtered flow quantities are zero outside $\Omega_{\mathrm{f}}$ beyond a distance $\delta$ from $\partial\Omega_{\mathrm{f}}$.}
    \label{fig:sketchdomain}
\end{figure}

In LES, the flow quantities $\varPhi$, i.e., the velocity and pressure fields, are assumed to be filtered by the convolution operation (see, e.g., \citet{Sagaut2005})
\begin{align}
\label{eq:LESfiltering}
    \widetilde{\varPhi}(\boldsymbol{x}) = \int\displaylimits_{\Omega_{\infty}} \varPhi(\boldsymbol{y})g(|\boldsymbol{x}-\boldsymbol{y}|) \mathrm{d}V_y, 
\end{align}
where $\Omega_{\infty}$ is an infinitely large three-dimensional domain, $\widetilde{\varPhi}$ is the filtered flow quantity, and $g$ is the filter kernel that satisfies
\begin{align}
\label{eq:integralfilterkernel}
    \int \displaylimits_{\Omega_{\infty}} g(\boldsymbol{|x|})\mathrm{d}V_x = 1.
\end{align}
The filtering operation removes the small flow scales and allows to formulate the equations governing large scales of the flow, the FNSE, and solve them with a coarser fluid mesh resolution than the resolution necessary to resolve the flow down to the Kolmogorov scales. \\
A fundamental limitation of the LES filtering operation occurs when filtering is performed close to the boundary of the flow domain, such as close to solid walls. The flow quantities have to be defined everywhere in $\Omega_{\infty}$, or the filtering operation has to be restricted to a domain with a minimal distance equal to the support of $g$ from walls. Note that with boundary of the flow domain, we refer to the boundary of the physical flow domain. Inlets and outlets in simulations are not boundaries of the physical domain but numerical interventions to restrict the size of the numerical domain. Filtered flow quantities are well defined near inlets or outlets, which is why such numerical boundaries are not addressed in the theory part of the present paper. Their numerical treatment, however, is discussed in section \ref{ssec:inletoutlet}. \\
When volume-filtering the flow, first introduced by \citet{Anderson1967} in the context of particle-laden flows, the convolution integral in equation \eqref{eq:LESfiltering} is only an integral over the flow domain $\Omega_{\mathrm{f}} \subset \Omega_{\infty}$, the domain where flow quantities are defined. Equivalently, the volume-filtering can also be an integral over $\Omega_{\infty}$ when the integrand is weighted with the fluid indicator function, $I_{\mathrm{f}}$, which is defined as
\begin{align}
    I_\mathrm{f}(\boldsymbol{x})=\begin{cases}
        1 & \text{if } \boldsymbol{x} \in \Omega_\mathrm{f},\\
        0 & \text{else}.
    \end{cases}
\end{align}
Therefore, the volume-filtered flow quantity, $\epsilon_{\mathrm{f}}(\boldsymbol{x}) \overline{\varPhi}(\boldsymbol{x})$, is defined as
\begin{align}
\label{eq:volumefiltering}
    \epsilon_{\mathrm{f}}(\boldsymbol{x}) \overline{\varPhi}(\boldsymbol{x}) = \int\displaylimits_{\Omega_{\mathrm{f}}}\varPhi(\boldsymbol{y})g(|\boldsymbol{x}-\boldsymbol{y}|) \mathrm{d}V_y = \int\displaylimits_{\Omega_{\infty}} I_{\mathrm{f}}(\boldsymbol{y}) \varPhi(\boldsymbol{y})g(|\boldsymbol{x}-\boldsymbol{y}|) \mathrm{d}V_y, 
\end{align}
where $\epsilon_{\mathrm{f}}$ is the local fluid volume fraction and given as 
\begin{align}
    \epsilon_{\mathrm{f}}(\boldsymbol{x}) = \int\displaylimits_{\Omega_{\mathrm{f}}}g(|\boldsymbol{x}-\boldsymbol{y}|) \mathrm{d}V_y  = \int\displaylimits_{\Omega_{\infty}} I_{\mathrm{f}}(\boldsymbol{y})g(|\boldsymbol{x}-\boldsymbol{y}|) \mathrm{d}V_y.
\end{align}
The volume-filtered quantities are defined everywhere in $\Omega_{\infty}$, even outside the flow domain, but volume-filtered quantities are zero at distances from the flow domain that are larger than the support of $g$, which is referred to as $\delta$. Temporal dependencies are omitted for conciseness. Figure \ref{fig:sketchdomain} illustrates the effect of volume-filtering a confined flow domain, $\Omega_{\mathrm{f}}$, that is bounded by $\partial\Omega_{\mathrm{f}}$. The volume-filtered domain is not bounded and volume-filtered flow quantities are defined everywhere in $\Omega_{\infty}$. Beyond a distance $\delta$ from $\partial\Omega_{\mathrm{f}}$, which equals the support of the filter kernel, volume-filtered flow quantities are zero.  \\
Assuming a symmetric and spatially uniform filter kernel $g$, it is shown that applying the volume-filtering operation to the Navier-Stokes equations (NSE) gives (see \citet{Hausmann2024a})
\begin{align}
\label{eq:reducedNSEconstinuity}
    \dfrac{\partial \epsilon_{\mathrm{f}}}{ \partial t} + \dfrac{\partial u_{\epsilon,i}}{\partial x_i} &= 0, \\
\label{eq:reducedNSEmomentum}
    \rho_{\mathrm{f}}\dfrac{\partial u_{\epsilon,i}}{\partial t} + \rho_{\mathrm{f}}\dfrac{\partial}{\partial x_j}(u_{\epsilon,i}u_{\epsilon,j}) &= -\dfrac{\partial p_{\epsilon}}{\partial x_i} +\mu_{\mathrm{f}} \dfrac{\partial^2u_{\epsilon,i}}{\partial x_j \partial x_j} -s_i\color{black} + \mu_{\mathrm{f}} \mathcal{E}_i- \rho_{\mathrm{f}}\dfrac{\partial }{\partial x_j}\tau_{\mathrm{sfs},ij},
\end{align}
where $u_{\epsilon,i}=\epsilon_{\mathrm{f}}\overline{u}_i$ is the volume-filtered velocity and $p_{\epsilon}=\epsilon_{\mathrm{f}}\overline{p}$ is the volume-filtered pressure. Three closure terms arise in the volume-filtered momentum equation that are briefly discussed in the following. Note that the volume-filtered NSE as they are written in equations \eqref{eq:reducedNSEconstinuity}-\eqref{eq:reducedNSEmomentum} can be solved with a classical single-phase flow solver. \\
The first closure term is the momentum source, $s_i$, which arises from switching filtering and spatial derivative in the pressure term and the viscous term in the volume-filtered NSE and is given as
\begin{align}
\label{eq:momentumsource}
    s_i(\boldsymbol{x}) = \int\displaylimits_{\partial\Omega_{\mathrm{f}}}g(|\boldsymbol{x}-\boldsymbol{y}|)\left(-p \delta_{ij} + \mu_{\mathrm{f}} \left(\dfrac{\partial u_i}{\partial y_j}+\dfrac{\partial u_j}{\partial y_i}\right)\right)n_j\mathrm{d}A_y,
\end{align}
where the normal vector, $n_j$, points into the fluid. As illustrated in figure \ref{fig:sketchdomain}, $s_i$ is only non-zero in a region with the thickness $2\delta$ near the walls of the physical flow configuration. At sufficient distance from the wall, volume-filtering and spatial derivative commute for uniform filters and the momentum source term is zero.\\
The second closure term, the viscous closure $\mathcal{E}_i$, arises from switching filtering and spatial derivative a second time in the viscous term. The third closure term, the subfilter stress tensor, arises from filtering the non-linear advective term in the fluid momentum equation and is defined as
\begin{align}
\label{eq:definitiontau}
    \tau_{\mathrm{sfs},ij} = \epsilon_{\mathrm{f}} \overline{u_i u_j} - \epsilon_{\mathrm{f}}\Bar{u}_i \epsilon_{\mathrm{f}}\Bar{u}_j,
\end{align}
whereas for a classical LES in an unbounded domain, the subfilter stress closure reads
\begin{align}
    \tau_{\mathrm{sfs},ij}^{\mathrm{class.}} = \widetilde{u_i u_j} - \Tilde{u}_i \Tilde{u}_j.
\end{align}
Inside $\Omega_{\mathrm{f}}$ at a distance from $\partial\Omega_{\mathrm{f}}$ larger than $\delta$, the FNSE are recovered since $\epsilon_{\mathrm{f}}=1$, $s_i=0$, and $\mathcal{E}_i=0$. Inside the domain and far away from the boundary, the subfilter stress tensor equals the closure term known from classical LES. A comprehensive discussion of the three closures can be found in \citet{Hausmann2024a}. \\
Although many different filter kernel functions are symmetric, uniform, and satisfy equation \eqref{eq:integralfilterkernel}, for the remainder of this paper, it is assumed that the filter kernel is a Gaussian with standard deviation $\sigma$, i.e.,
\begin{align}
    g(\boldsymbol{x}) = \dfrac{1}{(2\pi \sigma^2)^{3/2}}\exp\left( -\dfrac{|\boldsymbol{x}|^2}{2 \sigma^2} \right).
\end{align}
The discussions on volume-filtering, however, hold for any symmetric and uniform filter kernel that satisfies equation \eqref{eq:integralfilterkernel}. \\
Note that when referring to the support of the Gaussian filter kernel, we refer to the the region where the Gaussian significantly deviates from zero ($\delta=4\sigma$ in the present study). Furthermore, we refer to the standard deviation of the Gaussian filter kernel, $\sigma$, as the filter width.

\subsection{Volume-filtering of wall-bounded flows}
\begin{figure}
    \centering
    \includegraphics[scale=1]{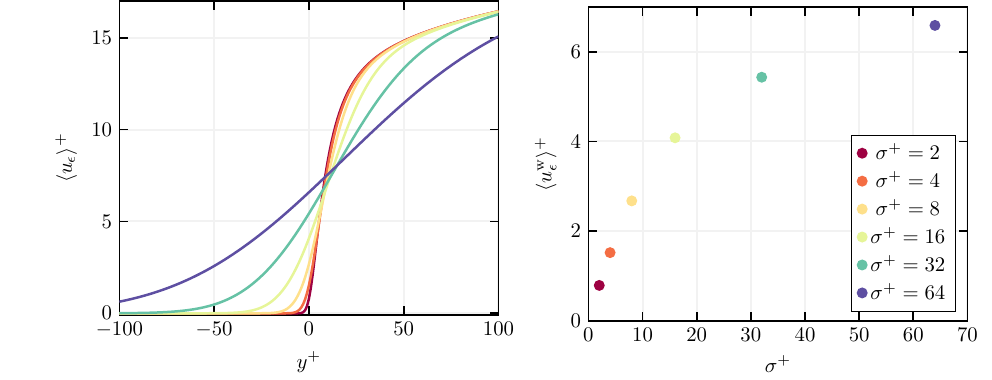}
    \caption{Explicitly volume-filtered mean streamwise velocity as a function of the wall normal coordinate $y$ (left) and evaluated at the wall (right) of a turbulent channel flow at $\mathrm{Re}_{\tau}=1000$. The $+$ indicates dimensionless quantities in wall units.}
    \label{fig:filteredvelocityprofiles}
\end{figure}

We assume that the walls do not move in the considered frame of reference, i.e., the fluid volume fraction does not change in time. This leads to the following simplifications in the volume-filtered NSE (see \citet{Hausmann2024a}):
\begin{align}
    \dfrac{\partial \epsilon_{\mathrm{f}}}{ \partial t} = 0, \quad \mathcal{E}_i = 0.
\end{align}
Therefore, the governing equations for the volume-filtered velocity and pressure in the case of fixed walls are 
\begin{align}
\label{eq:reducedNSEconstinuitywall}
    \dfrac{\partial u_{\epsilon,i}}{\partial x_i} &= 0, \\
\label{eq:reducedNSEmomentumwall}
    \rho_{\mathrm{f}}\dfrac{\partial u_{\epsilon,i}}{\partial t} + \rho_{\mathrm{f}}\dfrac{\partial}{\partial x_j}(u_{\epsilon,i}u_{\epsilon,j}) &= -\dfrac{\partial p_{\epsilon}}{\partial x_i} +\mu_{\mathrm{f}} \dfrac{\partial^2u_{\epsilon,i}}{\partial x_j \partial x_j} -s_i\color{black} - \rho_{\mathrm{f}}\dfrac{\partial }{\partial x_j}\tau_{\mathrm{sfs},ij}.
\end{align}
At a distance from the wall that is larger than $\delta$, the volume-filtered velocity is identical to the definition of the filtered velocity in classical LES. Volume-filtering near the wall leads to well-defined volume-filtered velocity profiles, as it is illustrated in figure \ref{fig:filteredvelocityprofiles} with the mean velocity profile of a turbulent channel flow. According to the no-slip boundary condition, the mean streamwise velocity, $\langle u \rangle$, at the wall is zero for the unfiltered case. For finite filter widths, there is a non-zero mean volume-filtered velocity at the wall, $\langle u_{\epsilon}^{\mathrm{w}} \rangle^+$, and a non-zero volume-filtered velocity outside the flow domain, which converges to zero with increasing distance from the flow domain outside of the flow domain. \\
In general, there is also a non-zero wall-normal velocity, $v_{\epsilon}$, at the wall. Despite this flux of the volume-filtered velocity through the physical flow domain boundaries, the total fluid mass is conserved. This becomes clear when recalling that the volume-filtered velocity is zero outside the flow domain beyond a distance $\delta$ from $\partial\Omega_{\mathrm{f}}$. This means that the flux of mass, momentum, energy, and any higher moments through the imaginary boundary with the distance $\delta$ from $\partial\Omega_{\mathrm{f}}$ is zero.\\
As shown by \citet{Bae2018}, a no-slip and no-penetration boundary condition in combination with a coarse near-wall fluid mesh resolution causes significant overprediction of the temporal fluctuations of the streamwise velocity. It is further shown, that by applying a non-zero slip and transpiration velocity (non-zero wall-tangential and wall-normal velocity) at the boundary instead, unphysically long streamwise velocity streaks are broken up and more realistic streamwise velocity fluctuations are observed. Volume-filtering, a natural extension of filtering to near-wall regions, provides the missing mathematical explanation directly derived from the NSE for the necessity of slip and transpiration boundary conditions for coarse near-wall fluid mesh resolutions.

\section{Wall-modeling derived from volume-filtering}
\label{sec:modeling}
In the present section, an alternative to classical wall-modeling is derived from the volume-filtered equations \eqref{eq:reducedNSEconstinuitywall}-\eqref{eq:reducedNSEmomentumwall}. Since volume-filtered quantities are defined in an infinite domain, there are no boundaries for volume-filtered quantities. In practice, however, homogeneous Dirichlet boundary conditions can be applied for the volume-filtered velocity at a distance larger $\delta$ from the flow domain in non-periodic directions, since the velocity is by definition zero there. Therefore, wall-modeling in the context of volume-filtering means primarily modeling of the momentum source $s_i$ instead of modeling the boundary conditions as in classical wall models. In addition, the subfilter stress tensor, $\tau_{\mathrm{sfs},ij}$, requires modeling, such as in every LES application. The modeling that we propose for the momentum source and the modeling used for the subfilter stress tensor is described in the following.

\subsection{Modeling $s_i$ with the PC-IBM}
\label{ssec:modelingsi}
The momentum source term $s_i$ contains the unfiltered pressure and viscous stresses on the wall. Since these stresses are convoluted with the filter kernel $g$ over the wall (see equation \eqref{eq:momentumsource}), the momentum source is distributed within a thickness equal to the support of $g$ around the wall, which is illustrated in figure \ref{fig:sketchdomain}. Note that the integral of $s_i$ over $\Omega_{\infty}$ equals the total hydrodynamical force, $\boldsymbol{F}_{\mathrm{h}}$, acting on the walls, i.e.,
\begin{align}
    \int\displaylimits_{\Omega_{\infty}} s_i(\boldsymbol{x}) \mathrm{d}V_{x} = F_{\mathrm{h},i}.
\end{align}
Since the unfiltered pressure and viscous stresses are not available in the volume-filtered framework, the momentum source term, $s_i$, requires modeling. We model $s_i$ similar as in the recently proposed PC-IBM \cite{Hausmann2024b}. For the sake of conciseness, we abbreviate the fluid stress vector as 
\begin{align}
    \Sigma_i = \left(-p \delta_{ij} + \mu_{\mathrm{f}} \left(\dfrac{\partial u_i}{\partial y_j}+\dfrac{\partial u_j}{\partial y_i}\right)\right)n_j,
\end{align}
such that $s_i$ is given as 
\begin{align}
    s_i(\boldsymbol{x}) = \int\displaylimits_{\partial\Omega_{\mathrm{f}}}g(|\boldsymbol{x}-\boldsymbol{y}|)\Sigma_i(\boldsymbol{y})\mathrm{d}A_y.
\end{align}
By discretizing the wall in $N_{\mathrm{s}}$ surface elements, the momentum source can be approximated as 
\begin{align}
\label{eq:particlemomentumsourcediscrete}
     s_i(\boldsymbol{x}) \approx \sum_{l=1}^{N_{\mathrm{s}}} g(|\boldsymbol{x}-\boldsymbol{X}_l|)\Sigma_{l,i} \Delta A_l,
\end{align}
where $\boldsymbol{X}_l$ is the position of the center of a surface element, which we refer to as surface marker, $\Delta A_l$ is the area of a surface element, and $\Sigma_{l,i}$ is the yet unknown fluid stress vector associated with the surface element with the index $l$. A surface element can be, for instance, one triangle of a triangulated surface with the center $\boldsymbol{X}_l$ and the area $\Delta A_l$. \\
Since the volume-filtered velocity at the wall can be estimated, which is described in the subsequent section, the fluid stress vector at each surface element can be determined, such that this desired volume-filtered velocity at the wall is achieved. This is the essential concept of the PC-IBM that unifies the volume-filtering framework with the established numerical methodology of the IBM in a physically consistent manner \cite{Hausmann2024b}. \\
On a Cartesian fluid mesh, the surface markers generally do not coincide with the collocation points of the fluid mesh. The flow quantities are interpolated from the fluid mesh to the surface markers with the same Gaussian that is assumed as filter kernel, which means that the interpolation essentially constitutes another filtering operation of the already filtered flow quantities on the fluid mesh. Filtering twice with a Gaussian filter kernel with a filter width $\sigma$ is equivalent to filtering once with $\sqrt{2}\sigma$ \cite{Hausmann2024a}. Therefore, all quantities known at the surface marker on the wall possess a filter width $\sigma_{\mathrm{w}}=\sqrt{2}\sigma$, which is the filter width of the corresponding filter kernel $g_{\mathrm{w}}$. \\
In practice, the fluid stress vector is computed as
\begin{align}
\label{eq:computationstressvector}
    \boldsymbol{\Sigma}_{l} \approx \boldsymbol{b}_l \mathcal{L}(\boldsymbol{X}_l),
\end{align}
where the length scale, $\mathcal{L}(\boldsymbol{X}_l)$, for the surface marker with the index $l$ is given as
\begin{align}
\label{eq:lengthscale}
    \mathcal{L}(\mathbf{X}_l) = \dfrac{1}{\sum_{m=1}^{N_{\mathrm{s}}} g_{\mathrm{w}}(|\boldsymbol{X}_l-\boldsymbol{X}_m|) \Delta A_m},
\end{align}
and the vector $\boldsymbol{b}_l$ contains the advective term, $\mathcal{A}_i^n$, the pressure term $\mathcal{P}_i^n$, the viscous term, $\mathcal{V}_i^n$, and the transient term of the volume-filtered momentum equation interpolated to the surface marker with the index $l$. After discretization of the transient term with finite differences, the vector $\boldsymbol{b}_l$ is given as
\begin{align}
\label{eq:vectorb}
    b_{l,i} = -\rho_{\mathrm{f}}\dfrac{u_{\epsilon,i}^{\mathrm{des}}(\boldsymbol{X}_l) - u_{\epsilon,i}^n(\boldsymbol{X}_l)}{\Delta t} -\mathcal{A}_i^n(\boldsymbol{X}_l) - \mathcal{P}_i^n(\boldsymbol{X}_l) + \mathcal{V}_i^n(\boldsymbol{X}_l).
\end{align}
The superscript $n$ refers to the previous time level. The interpolated quantities and the desired volume-filtered velocity, $u_{\epsilon,i}^{\mathrm{des}}$, possess a filter width $\sigma_{\mathrm{w}}$. The derivation and implementation details of the PC-IBM may be found in \citet{Hausmann2024b}.

\subsection{Obtaining the volume-filtered velocity at the wall}
\label{ssec:volumefilteredvelocityatwall}
\begin{figure}
    \centering
    \includegraphics[scale=0.9]{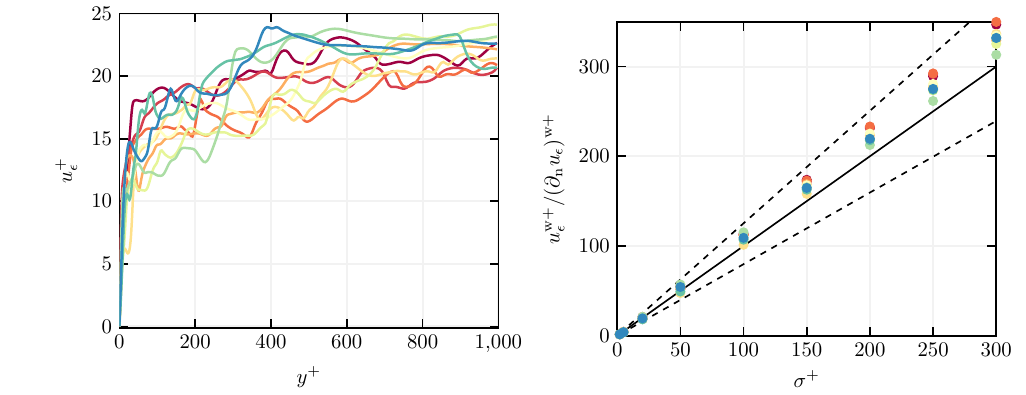}
    \caption{Randomly picked instantaneous streamwise velocity profiles of a turbulent channel flow with $\mathrm{Re}_{\lambda}=1000$ from the Johns Hopkins turbulence database \cite{Graham2017} (left) as a function of the wall distance. The ratio between the volume-filtered velocity and the wall-normal gradient of the volume-filtered velocity is plotted over the filter width (right). The upper dashed line corresponds to $\sqrt{\dfrac{\pi}{2}}\sigma^+$, the lower dashed line to $\sqrt{\dfrac{2}{\pi}}\sigma^+$, and the solid line to $\sigma^+$. }
    \label{fig:instantaneousfilteredvelocity}
\end{figure}

An essential part of the proposed wall-modeling is the estimation of the desired volume-filtered velocity at the wall, since it directly influences the momentum source and, therefore, the resulting flow field. A perfect model for the desired volume-filtered velocity would provide the explicitly volume-filtered velocity at the wall, as it is demonstrated in figure \ref{fig:filteredvelocityprofiles} with the mean streamwise velocity of a channel flow. Since the exact unfiltered flow field near the wall, as it is obtained from a DNS, is not known in the volume-filtering framework, the explicit volume-filtering operation cannot be performed and the volume-filtered velocity at the wall requires modeling. \\
In the originally proposed PC-IBM \cite{Hausmann2024b}, the volume-filtered velocity at the wall is estimated with the approximate deconvolution method \cite{VanCittert1931,Stolz1999}, which has been shown to work well for the flow around particles with relatively small Reynolds numbers relative to the large Reynolds numbers of the turbulent flows considered in the present study. The idea of the approximate deconvolution method is to compute the volume-filtered velocity at the wall for filter widths larger than $\sigma_{\mathrm{w}}$, which are used to extrapolate the volume-filtered velocity to $\sigma=0$. The resulting estimate of the unfiltered velocity can be used to estimate the correct volume-filtered velocity at the wall. The extrapolation, however, is inaccurate for large $\sigma$ because the volume-filtered velocity at the wall is very sensitive to $\sigma$, especially for small $\sigma$, as is evident from figure \ref{fig:filteredvelocityprofiles}. In wall-modeled LES, the fluid-mesh spacing in the wall-normal direction is typically chosen $\Delta y^+\geq \mathcal{O}(10)$. Assuming the filter width to be equal to the wall-normal fluid mesh spacing, this yields $\sigma^+\geq \mathcal{O}(10)$. \\
In the present paper, we propose a different method to estimate the volume-filtered fluid velocity at the wall, which is particularly suitable for large Reynolds number flows and large filter widths. A fundamental assumption for the proposed estimation is that the filtering operation of the near wall velocity field is mainly a filtering in the wall-normal direction, $x_{\mathrm{n}}$, which may be justified by the fact that the velocity varies much steeper in the wall-normal direction than in the wall-tangential directions, $x_{\mathrm{s}}$ and $x_{\mathrm{t}}$. Mathematically, this assumption is expressed as 
\begin{align}
    u_{\epsilon,i}^{\mathrm{w}} = u_{\epsilon,i}(x_{\mathrm{s}},x_{\mathrm{t}},x_{\mathrm{n}}=0)\approx \int\displaylimits_0^{\infty}g_{1\mathrm{D}}(0-\Tilde{x}_{\mathrm{n}})u_i(x_{\mathrm{s}},x_{\mathrm{t}},\Tilde{x}_{\mathrm{n}}) \mathrm{d}\Tilde{x}_{\mathrm{n}},
\end{align}
where the one-dimensional Gaussian is given as
\begin{align}
    g_{1\mathrm{D}}(x) = \dfrac{1}{\sigma \sqrt{2\pi}}\exp{\left(-\dfrac{x^2}{2\sigma^2}\right)}.
\end{align}
Figure \ref{fig:instantaneousfilteredvelocity} shows several randomly picked instantaneous streamwise velocity profiles of a turbulent channel flow from the Johns Hopkins turbulence database \cite{Graham2017} with a friction velocity Reynolds number $\mathrm{Re}_{\tau}=u_{\tau}h/\nu=1000$ as a function of the distance to the wall, $y$, where $u_{\tau}=\sqrt{\tau_{\mathrm{w}}/\rho_{\mathrm{f}}}$ is the friction velocity, $l_{\tau}=\nu/u_{\tau}$ is the viscous length scale, $\tau_{\mathrm{w}}$ is the mean wall shear stress, and $h$ is the channel half-height. Quantities with the superscript $+$ are normalized with the wall units $u_{\tau}$ and $l_{\tau}$. After volume-filtering the streamwise velocity at the wall and dividing by the wall-normal gradient of the volume-filtered velocity at the wall, $(\partial_nu_{\epsilon,i})^{\mathrm{w}}$, a surprisingly simple relation to the filter width is revealed. In figure \ref{fig:instantaneousfilteredvelocity} it is observed that $\dfrac{u_{\epsilon,\alpha}^{\mathrm{w}}}{(\partial_nu_{\epsilon,\alpha})^{\mathrm{w}}}\approx \sigma$ holds with good accuracy even for large filter widths. Note that no summation is carried out over the index $\alpha$. This relation may be explained with the similarity of the streamwise velocity profile to root-functions. The idea of approximating the streamwise velocity profile with root-functions is not new. In the wall model of \citet{Werner1993}, the mean streamwise velocity profile is assumed to be proportional to $x_{\mathrm{n}}^{1/7}$. We make a more general approach by assuming that the velocity profile may be expanded as infinite series of root-functions of arbitrary order, i.e.,
\begin{align}
\label{eq:rootexpansion}
    u(x_{\mathrm{n}}) = \sum_{k=1}^{\infty}a_k x_{\mathrm{n}}^{1/k},
\end{align}
where $a_k$ are real coefficients with the same sign, i.e., $\forall a_k \in \mathbb{R}, a_k\ge0$ or $\forall a_k \in \mathbb{R}, a_k\le0$. Note that this is a much weaker assumption than the commonly assumed log-law, which applies only to the mean velocity along planar walls with zero pressure gradient. For equation \eqref{eq:rootexpansion}, it is shown in the Appendix \ref{ap:limitsVTGR} that the ratio $\dfrac{u_{\epsilon,\alpha}^{\mathrm{w}}}{(\partial_nu_{\epsilon,\alpha})^{\mathrm{w}}}$ lies always within the limits
\begin{align}
    \sqrt{\dfrac{2}{\pi}}\sigma\le \dfrac{u_{\epsilon,\alpha}^{\mathrm{w}}}{(\partial_nu_{\epsilon,\alpha})^{\mathrm{w}}} \le \sqrt{\dfrac{\pi}{2}}\sigma.
\end{align}

This means that if the unfiltered velocity can be represented by an infinite series of root-functions in the wall-normal direction (at least in the close vicinity of the wall), $\dfrac{u_{\epsilon,\alpha}^{\mathrm{w}}}{(\partial_nu_{\epsilon,\alpha})^{\mathrm{w}}}=\sigma$ is true with a maximum error of approximately $25\%$. \\
It should be noted that in the case of small magnitudes of $(\partial_nu_{\epsilon,i})^{\mathrm{w}}$, which is typical for the wall-normal velocity and the wall-tangential velocity at the detachment and reattachment points in separated flows, the relation $\dfrac{u_{\epsilon,\alpha}^{\mathrm{w}}}{(\partial_nu_{\epsilon,\alpha})^{\mathrm{w}}}=\sigma$ is usually not a good approximation. Profiles with a small magnitude of  $(\partial_nu_{\epsilon,i})^{\mathrm{w}}$ cannot be approximated well with root-functions with coefficients of the same sign. However, a small wall-normal gradient leads to a volume-filtered velocity at the wall that is close to zero. Therefore, the absolute error of the desired volume-filtered velocity at the wall is still small. \\ 
Inspired by the observation that $\dfrac{u_{\epsilon,\alpha}^{\mathrm{w}}}{(\partial_nu_{\epsilon,\alpha})^{\mathrm{w}}}=\sigma$ is a good approximation, we propose to model the desired volume-filtered velocity at the wall as 
\begin{align}
\label{eq:modelwallvelocity}
    u_{\epsilon,i}^{\mathrm{des}}(\boldsymbol{X}_l) = \sigma_{\mathrm{w}}(\partial_nu_{\epsilon,i})^{\mathrm{w}}(\boldsymbol{X}_l),
\end{align}
where the wall-normal gradient of the volume-filtered velocity is evaluated by interpolating the gradient from the fluid mesh to the surface marker. The model assumes that if the velocity at the boundary, $u_{\epsilon,i}^{\mathrm{des}}(\boldsymbol{X}_l)$, is enforced correctly, the volume-filtered NSE produce the correct wall-normal gradient of the volume-filtered velocity at the wall. \\
Estimating the volume-filtered velocity at the wall using equation \eqref{eq:modelwallvelocity} is numerically stable, because it avoids unphysically large or small values of the volume-filtered velocity at the wall. If $u_{\epsilon,i}^{\mathrm{des}}(\boldsymbol{X}_l)$ is too large, i.e., $\dfrac{u_{\epsilon,\alpha}^{\mathrm{w}}}{(\partial_nu_{\epsilon,\alpha})^{\mathrm{w}}}>\sigma$, it will be decreased according to equation \eqref{eq:modelwallvelocity}, which simultaneously increases the gradient. This prevents the volume-filtered velocity at the wall from being further reduced in the next time step. The model is also very simple to implement and computationally efficient, because it does not require explicit filtering such as the approximate deconvolution method proposed in the original PC-IBM \cite{Hausmann2024b}.

\subsection{Required modifications of a classical IBM}
The proposed wall modeling can be realized with any standard IBM implementation after four modifications are implemented. Note that this assumes that the fluid volume fraction is not changing in time, i.e., that the wall does not move in the considered frame of reference. The following modifications are required:
\begin{itemize}
    \item The interpolation of fluid quantities to the surface markers must be carried out with a Gaussian interpolation kernel with a standard deviation $\sigma$. In practice, the Gaussian is truncated after a distance of $4\sigma$ from its center.
    \item The spreading of the IBM source, i.e., the computation of the discrete momentum source term $s_i$ given in equation \eqref{eq:particlemomentumsourcediscrete}, must be carried out with a Gaussian kernel with a standard deviation $\sigma$.
    \item The desired volume-filtered velocity is computed based on the wall-normal gradient of the volume-filtered velocity according to $u_{\epsilon,i}^{\mathrm{des}} = \sqrt{2}\sigma(\partial_nu_{\epsilon,i})^{\mathrm{w}}$.
    \item The IBM force contribution onto the wall associated to the surface marker $l$ is given as $\boldsymbol{F}_{\mathrm{IBM},l} = \boldsymbol{\Sigma}_{l}\Delta A_l = \boldsymbol{b}_l \mathcal{L}(\boldsymbol{X}_l)\Delta A_l$ (no summation over $l$ here), where the length scale $\mathcal{L}$ is given in equation \eqref{eq:lengthscale} and the vector $\boldsymbol{b}_l$ is given in equation \eqref{eq:vectorb}.
\end{itemize}

\subsection{Modeling of the subfilter stress tensor}
The focus of the present study is the wall-modeling in the context of volume-filtering and not to find the most suitable model for the subfilter stress tensor. Therefore, we perform the test cases with the existing models for the subfilter stress tensor described in the following, while being aware of the fact that there may exist models that lead to better results. \\
The subfilter stress tensor in the context of volume-filtering, $\tau_{\mathrm{sfs},ij}$, is closely related to its single-phase flow equivalent in classical LES, and even identical in regions in $\Omega_{\mathrm{f}}$ with at least a distance $\delta$ to the boundary, inside the flow domain. As seen from equation \eqref{eq:definitiontau}, $\tau_{\mathrm{sfs},ij}$ contains the fluid volume fraction, which deviates from $\epsilon_{\mathrm{f}} =1$ in the close vicinity of the wall. \\
\citet{Hausmann2024a} propose to model the subfilter stress tensor as (for fixed boundaries)
\begin{align}
    \tau_{\mathrm{sfs},ij}^{\mathrm{NL}} = \sigma^2\dfrac{\partial u_{\epsilon,i} }{\partial x_k}\dfrac{\partial u_{\epsilon,j} }{\partial x_k},
\end{align}
which is accurate up to terms of the order four in the filter width. Far away from the boundary, $\tau_{\mathrm{sfs},ij}^{\mathrm{NL}}$ converges to the non-linear model or non-linear gradient model \cite{Leonard1975,Liu1994b,Johnson2020a}, which is why we refer to this model as the non-linear model. The non-linear model is known for being highly correlated with the subfilter stress tensor but also to produce to few dissipation (see, e.g., \citet{Borue1998,Horiuti2003}), which is why it is often combined with a turbulent viscosity in so-called mixed models \cite{Vreman1996,Vreman1997}. \\
In addition to the non-linear model, we perform simulations with the Vreman model for the subfilter stress tensor adapted to the volume-filtering framework, which has been shown to perform well in WMLES of separated flow as compared to, e.g., the dynamic Smagorinsky model \cite{Zhou2024}. We adapt the Vreman model such that the turbulent viscosity $\nu_{\mathrm{t}}$ is computed from the volume-filtered velocity and we assume an isotropic filter kernel, even though the fluid mesh may be anisotropic. The filter width can be generally chosen independent of the fluid mesh, given that the filter width is large enough to avoid flow structures that cannot be resolved by the fluid mesh. The turbulent viscosity is modeled as 
\begin{align}
    \nu_{\mathrm{t}} = C_{\mathrm{Vre}}\sigma^2\sqrt{\dfrac{B_{\beta}}{\alpha_{ij}\alpha_{ij}}},
\end{align}
where the Vreman constant is chosen to have the value proposed by \citet{Vreman2004}, i.e., $C_{\mathrm{Vre}}=0.025$, and
\begin{align}
    \alpha_{ij} &= \dfrac{\partial u_{\epsilon,j}}{\partial x_i}, \\
    \beta_{ij} &= \alpha_{mi}\alpha_{mj}, \\
    B_{\beta} &= \beta_{11}\beta_{22} - \beta_{12}^2 + \beta_{11}\beta_{33} - \beta_{13}^2 + \beta_{22}\beta_{33} - \beta_{23}^2.
\end{align}
As a third model for the subfilter stress tensor, the non-linear model and the Vreman model are combined to a mixed model that is supposed to retain the high correlations of the non-linear model with the subfilter stress tensor while providing sufficient dissipation. With the mixed model, the subfilter stress tensor is approximated as
\begin{align}
    \tau_{\mathrm{sfs},ij}^{\mathrm{Mix}} = \sigma^2\dfrac{\partial u_{\epsilon,i} }{\partial x_k}\dfrac{\partial u_{\epsilon,j} }{\partial x_k} - 2\nu_{\mathrm{t}}\overline{S}_{ij},
\end{align}
where the volume-filtered strain-rate tensor is given as 
\begin{align}
    \Bar{S}_{ij} = \dfrac{1}{2}\left[ \dfrac{\partial u_{\epsilon,i}}{\partial x_j} + \dfrac{\partial u_{\epsilon,j}}{\partial x_i} \right].
\end{align}
The models for the subfilter stress tensor mentioned above are valid for boundaries that are not moving in the considered frame of reference. In the case of moving boundaries, additional terms appear to ensure Galilean invariance of the volume-filtered NSE (see \citet{Hausmann2024a} for details).

\subsection{Inlet and outlet boundary conditions}
\label{ssec:inletoutlet}
The problem of finding appropriate inlet and outlet boundary conditions with the VF-WMLES is similar to finding the inlet and outlet boundary conditions for classical LES (with modeled or resolved walls) or DNS. If a method exists to generate a turbulent inflow or outflow for a DNS, the generated field can be explicitly volume-filtered to obtain the volume-filtered inflow condition for the VF-WMLES. Outside the flow domain beyond a distance $\delta$ from the wall, the inlet and outlet boundary conditions must be consistent with the fact that flow quantities are zero in this region. Inside the flow domain beyond a distance $\delta$ from the wall, flow quantities in the VF-WMLES are identical to flow quantities in a classical LES and, therefore, the inlet and outlet boundary conditions are also the same as in a classical LES in this region. Regardless of whether a DNS, a classical LES, or a VF-WMLES is performed, inlet and outlet boundary conditions are generally approximations.

\section{Results and Discussions}
\label{sec:results}
In the present section, the proposed wall-modeling is validated with two common configurations, a turbulent channel flow and a periodic hill. In both configurations the model for the subfilter stress tensor and the spatial resolution is varied. The simulations are carried out with a finite volume flow solver that is second order in space and time and that relies on monolithic solution of the momentum and continuity equation using momentum weighted interpolation. Details on the flow solver and its implementation may be found in \citet{Denner2014a,Bartholomew2018,Denner2020}. In the present study, the fluid mesh is Cartesian.

\subsection{Turbulent channel flow}
VF-WMLES of a turbulent channel flow with a friction velocity Reynolds number $\mathrm{Re}_{\tau}=1000$ are performed, which is the same physical configuration as described in \citet{Graham2016}. The physical domain size is $L_x\times L_y\times L_z=8\pi h\times 2h \times 3\pi h$ and the Reynolds number based on the bulk velocity, $u_{\mathrm{b}}$, is $\mathrm{Re}_{\mathrm{b}}=40,000$. The flow domain is periodic in the $x$- and $z$-direction and no-slip walls are assumed at the upper and lower end of the $y$-direction for the unfiltered velocity. The flow is driven by a constant pressure gradient in the $x$-direction. \\
The VF-WMLES are performed with two different spatial resolutions, referred to as fine and coarse, and two different filter widths, which are chosen equal to the smallest fluid mesh spacing, respectively. The fluid mesh information of the fine and the coarse configuration are summarized in table \ref{tab:meshchannel}. The size of the numerical domain in the non-periodic direction, $\overline{L}_y$, is larger than the size of the physical domain in this direction because volume-filtered quantities can be non-zero outside the flow domain, but they are guaranteed to be zero outside the channel beyond a distance $\delta$ from the physical boundaries. The boundaries of the numerical domain have at least a distance $\delta$ from the physical boundaries, which ensures that the fluid quantities are zero at the boundary of the numerical domain. In the present study, homogeneous Dirichlet boundary conditions are applied for the velocity and homogeneous Neumann boundary conditions are applied for the wall normal pressure gradient at the non-periodic boundaries. In both mesh configurations, a CFL-number of $0.5$ is maintained and the SuperBee flux limiter is applied using a flux limiter to gradient ratio of $L=2$ \cite{Denner2015a}. The simulations are carried out for at least 20 flow through times based on the bulk velocity, $L_x/u_{\mathrm{b}}$, before statistics are evaluated. \\

\begin{figure}
    \centering
    \includegraphics[scale=0.3]{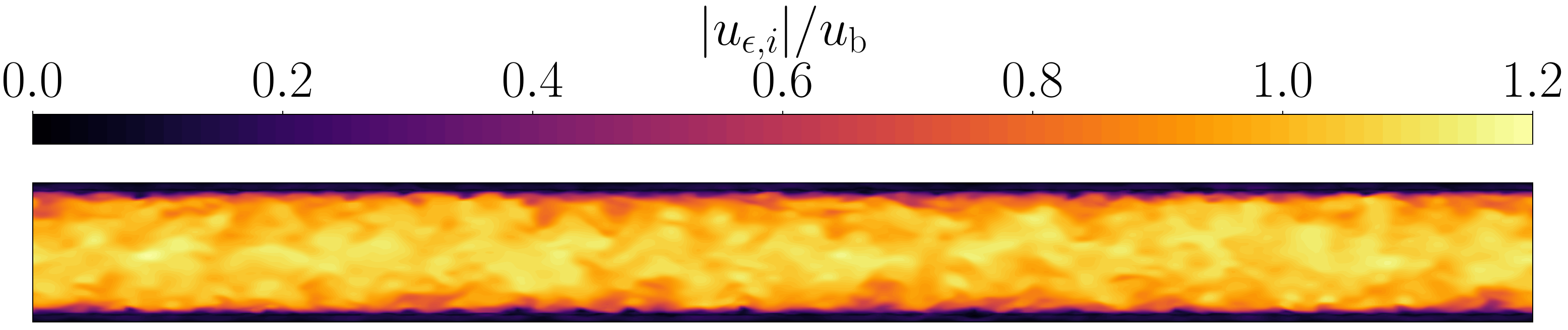}
    \caption{Snapshot of the absolute volume-filtered velocity of the turbulent channel flow obtained from the VF-WMLES with the fine fluid mesh configuration.}
    \label{fig:contourchannel}
\end{figure}

\begin{table}[htb]
\caption{\label{tab:meshchannel} Fluid mesh information and filter width for the fine and coarse configuration of the turbulent channel flow.}
\begin{ruledtabular}
\begin{tabular}{ccccc}
Name & $\overline{L}_y$ & $N_x\times N_y \times N_z$ & $\Delta x^+ \times \Delta y^+ \times \Delta z^+$ & $\sigma^+$\\
\hline
fine & $2.36h$ & $450 \times 80 \times 180$ & $55.8 \times 29.5 \times 52.3$ & $30.0$ \\
coarse & $2.7h$ & $251 \times 45 \times 94$ & $100.1\times 60.0 \times 100.2$ & $60.0$ \\
\end{tabular}
\end{ruledtabular}
\end{table}
Figure \ref{fig:contourchannel} shows a snapshot of the absolute volume-filtered velocity obtained from the VF-WMLES with the fine fluid mesh configuration. Inside the channel and far away from the walls, volume-filtering equals the classical LES filtering and the FNSE are solved similar as in a classical LES. At the wall, i.e., the boundary of the physical domain that lies inside the numerical domain, a non-zero velocity is observed, which rapidly converges to zero beyond the wall. At the boundary of the numerical domain, which lies outside the physical domain, the velocity is zero. \\

\begin{figure}[h]
    \centering
    \includegraphics[scale=0.9]{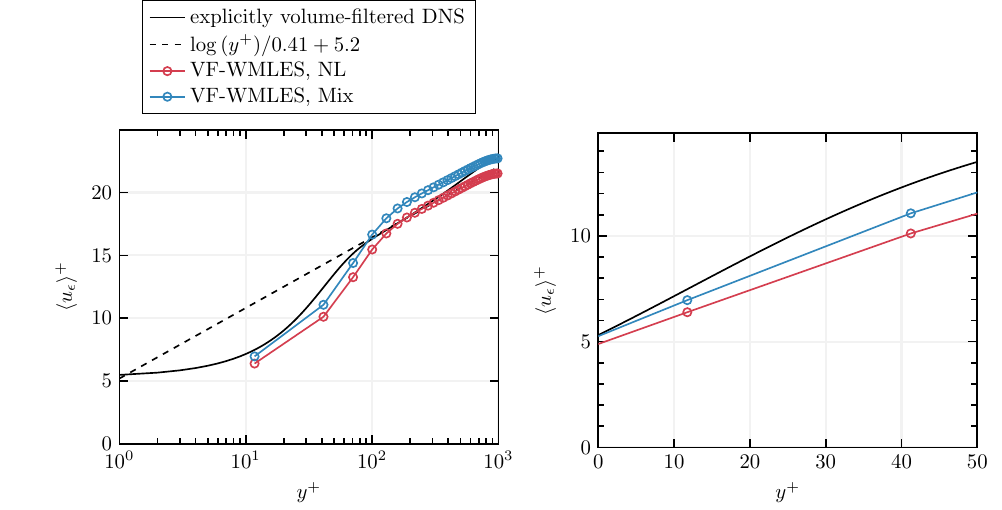}
    \includegraphics[scale=0.9]{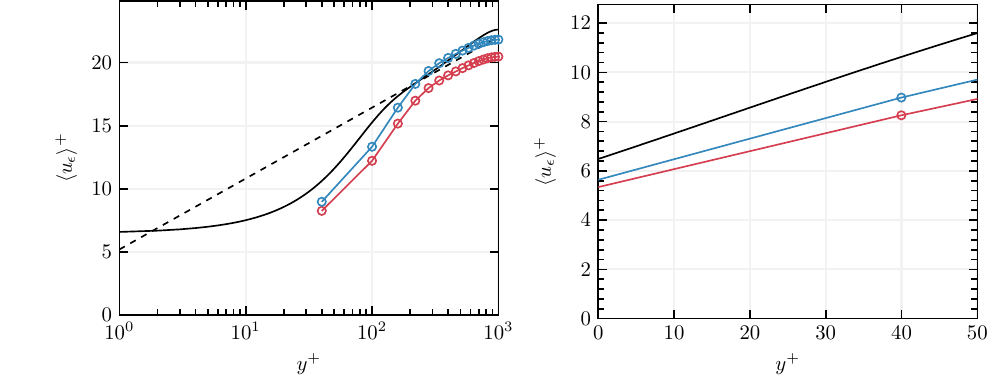}
    \caption{Mean profile of the streamwise velocity of the turbulent channel with logarithmic (left) and linear (right) scaling of the wall distance. The VF-WMLES results obtained with the non-linear model (NL) and the mixed model (Mix) are compared with the explicitly volume-filtered DNS from \citet{Graham2016}. The log-law is shown for comparison. The results at the top correspond to a filter width $\sigma^+=30$ and are obtained with the VF-WMLES with the fine fluid mesh, and the results at the bottom correspond to a filter width $\sigma^+=60$ and are obtained with the VF-WMLES with the coarse fluid mesh. }
    \label{fig:meanvelocitychannel}
\end{figure}

In figure \ref{fig:meanvelocitychannel}, the mean streamwise velocity of the turbulent channel flow obtained with the VF-WMLES and the fine fluid mesh configuration is compared to the explicitly volume-filtered DNS. The VF-WMLES is performed with the non-linear model and the mixed model. Note that the mean of a volume-filtered flow quantity equals the volume-filtered mean flow quantity, i.e.,
\begin{align}
    \dfrac{1}{T_1-T_0}\int\displaylimits_{T_0}^{T_1} \int\displaylimits_{\Omega_{\infty}} I_{\mathrm{f}}(\boldsymbol{y}) \varPhi(t,\boldsymbol{y})g(|\boldsymbol{x}-\boldsymbol{y}|) \mathrm{d}V_y \mathrm{d}t = \int\displaylimits_{\Omega_{\infty}} I_{\mathrm{f}}(\boldsymbol{y})g(|\boldsymbol{x}-\boldsymbol{y}|) \dfrac{1}{T_1-T_0}\int\displaylimits_{T_0}^{T_1}  \varPhi(t,\boldsymbol{y}) \mathrm{d}t \mathrm{d}V_y,
\end{align}
where $T_0$ and $T_1$ are the the lower and upper limits for the temporal averaging, respectively. The filter width for the explicit volume-filtering equals the filter width used in the VF-WMLES, i.e., $\sigma^+=30$. With a perfect model for the desired volume-filtered velocity at the wall, an exact formulation for the subfilter stress tensor, and without discretization errors, the VF-WMLES would recover the the explicitly volume-filtered DNS. However, a deviation is observed in figure \ref{fig:meanvelocitychannel} that is caused by modeling and discretization errors. The mean velocity profile is sensitive to the model for the subfilter stress tensor. The mixed model predicts a larger mean velocity than the non-linear model across the whole channel, but both models for the subfilter stress tensor recover the slope of the log-law. The wake region, i.e., the deviation of the velocity from the log-law in the center of the channel, is not predicted correctly. The most likely cause for that are modeling errors of the subfilter stress tensor. The mean volume-filtered velocity at the wall, however, is in very good agreement with the explicitly volume-filtered velocity at the wall. The mixed model predicts a slightly larger mean wall-normal velocity gradient at the wall than the non-linear model. Since according to the proposed model the volume-filtered velocity at the wall depends linearly on the volume-filtered wall-normal gradient at the wall, the mixed model predicts a slightly larger volume-filtered mean velocity at the wall than the non-linear model. The accurate prediction of the mean volume-filtered velocity at the wall suggests that the observed deviations inside the channel are likely to be caused by the model for the subfilter stress tensor rather than the newly proposed wall-model. This is supported by the findings of \citet{Bae2019}, who show that the shape of the mean velocity profile is sensitive to the model for the subfilter stress tensor. Different models for the subfilter stress tensor may improve the agreement of the volume-filtered velocity inside the channel, but the focus of the present work is the modeling of the slip and penetration velocity at the wall, which is why we do not perform a broad study of other models for the subfilter stress tensor. \\
The study of the mean streamwise velocity of the turbulent channel flow is repeated with the coarse fluid mesh configuration with a filter width twice as large as the filter width of the fine fluid mesh configuration. Figure \ref{fig:meanvelocitychannel} shows the resulting mean velocity profiles. It can be observed that at such a large filter width, the log-law is essentially not existent for the explicitly volume-filtered mean velocity. The volume-filtered mean velocities predicted by the WMLES are smaller than with the fine resolution and the mixed model leads to a larger volume-filtered mean velocity across the channel than the non-linear model. Even with the coarse resolution, the mean of the volume-filtered velocity at the wall is captured relatively accurate. The predictions of the volume-filtered mean velocity at the wall with both models for the subfilter stress tensor deviate less than 20\% from the explicitly volume-filtered velocity at the wall. Similar to the mean streamwise velocity inside the channel obtained with the fine resolution, the mean streamwise velocity obtained with the coarse resolution is very sensitive to the choice of the model for the subfilter stress tensor. 

\subsection{Turbulent flow over a periodic hill}
\begin{figure}
    \centering
    \includegraphics[scale=0.8]{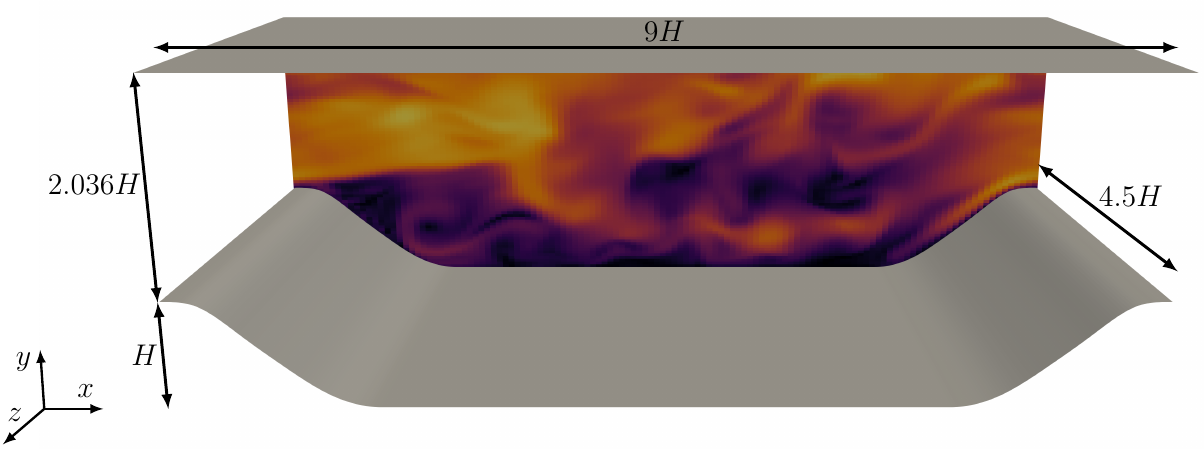}
    \caption{Sketch of the simulation configuration of the flow over periodic hills with a snapshot of the velocity field. The boundary at the top is a flat wall and the boundary at the bottom is a curved wall with a hill of the height $H$. The $x$-and $z$-direction is periodic and there is a mean flow driven in the positive $x$-direction. }
    \label{fig:sketchperiodichill}
\end{figure}

It is not sufficient for a wall model to perform well in turbulent flows over flat walls, since most of the applications of interest involve curved walls. In order to evaluate the VF-WMLES framework, and in particular the proposed modeling with the PC-IBM, in separated flows, the widely studied flow over periodic hills is simulated (see, e.g., \citet{Temmerman2003,Frohlich2005,Krank2018}). The simulation configuration is illustrated in figure \ref{fig:sketchperiodichill}. The physical domain is bounded by a box of the size $L_x \times L_y \times L_z= 9H \times 3.036H \times 4.5H$, where $H$ refers to the height of the hill. The flow is periodic in the $x$- and $z$-direction. The Reynolds number is $\mathrm{Re}_{\mathrm{H}}=\dfrac{u_{\mathrm{b}}H}{\nu_{\mathrm{f}}}=10,595$ with the the bulk velocity, $u_{\mathrm{b}}$, being defined as the volume flux divided by the cross sectional area of the channel at the crest of the hill. The desired volume flux, $Q_{\mathrm{des}}$, is enforced by applying a momentum source in the streamwise direction, $s_{\mathrm{drive}}$, which is adjusted every time step. This driving source is computed as described in \citet{Krank2018}, where the driving source of the next time step is given as 
\begin{align}
    s_{\mathrm{drive}}^{n+1} = s_{\mathrm{drive}}^{n} + C_{\mathrm{drive}}\left(\dfrac{Q_{\mathrm{des}}-Q^n}{\Delta t^n} - \dfrac{Q^n-Q^{n-1}}{\Delta t^{n-1}}\right),
\end{align}
where $s_{\mathrm{drive}}^{n}$, $Q^n$, and $\Delta t^n$ are the present driving source, volume flux, and time step, respectively. The superscript $n-1$ refers to quantities of the previous time step and $C_{\mathrm{drive}}$ is a constant which is empirically chosen to be $C_{\mathrm{drive}}=0.001$. This choice of $C_{\mathrm{drive}}$ enforces the desired volume flux with at least four digits of accuracy. \\
The simulations with the VF-WMLES are carried in a numerical domain that is extended in the $y$-direction dependent on the filter width to solve for the flow in the near wall region outside the physical domain. At both boundaries of the numerical domain of the non-periodic $y$-direction, which have at least a distance $\delta$ from the walls, homogeneous Dirichlet boundary conditions for the velocity and homogeneous Neumann conditions for the wall-normal pressure gradient are applied. The VF-WMLES is solved on a fine fluid mesh and a coarse fluid mesh with two different filter widths, each corresponding to the smallest fluid mesh spacing. The numerical domain size in $y$-direction, $\overline{L}_y$, the spatial resolution, and the filter width are summarized in table \ref{tab:meshperiodichill}. In the periodic hill simulations a van Leer flux limiter with a flux limiter to gradient ratio of $L=2$ is used \cite{Denner2015a} and the CFL-number is $0.1$. The simulations are carried out for at least 20 flow through times based on the bulk velocity at the crest of the hill, $L_x/u_{\mathrm{b}}$, before statistics are evaluated. \\

\begin{figure}
    \centering
    \includegraphics[scale=1]{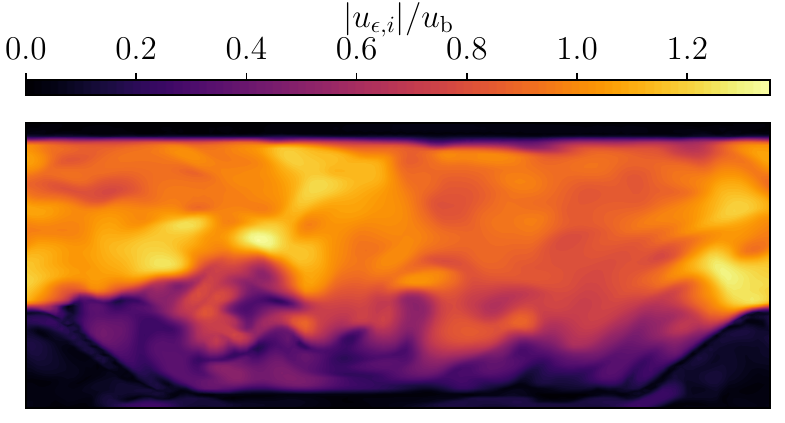}
    \caption{Snapshot of the absolute volume-filtered velocity of the periodic hill flow obtained from the VF-WMLES with the fine fluid mesh configuration.}
    \label{fig:contourperiodichill}
\end{figure}

\begin{figure}
    \centering
    \includegraphics[scale=0.9]{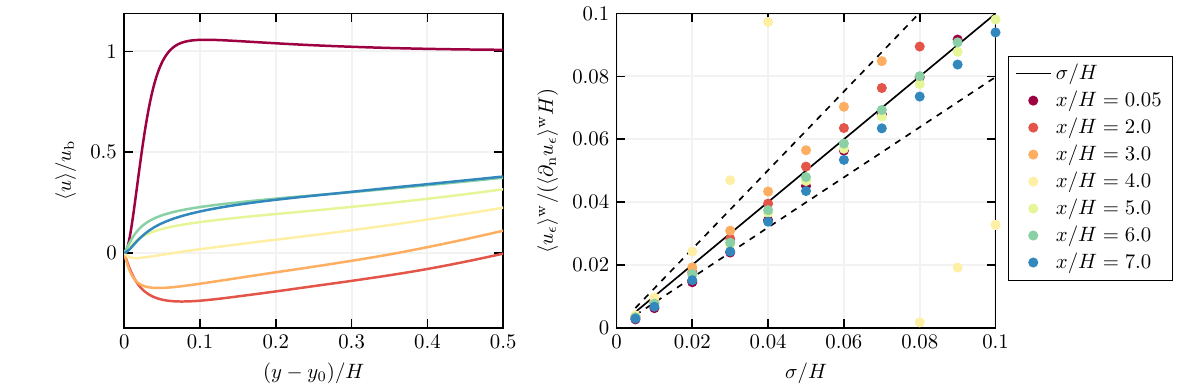}
    \caption{Mean streamwise DNS velocity profiles of the flow over periodic hills, which is taken from \citet{Krank2018}, as a function of the distance from the lower wall $y-y_0$ (left) and the corresponding ratio of the volume-filtered velocity at the wall and the volume-filtered wall-normal gradient for different filter widths (right). The profiles are shown for different positions in the streamwise direction $x/H$.}
    \label{fig:aprioriPeriodicHill}
\end{figure}

\begin{figure}
    \centering
    \includegraphics[scale=1.2]{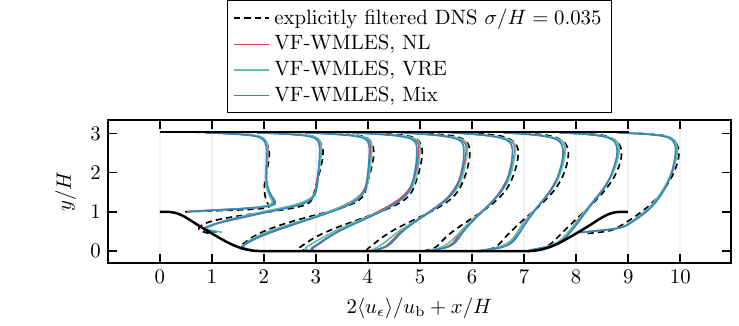}
    \includegraphics[scale=1.2]{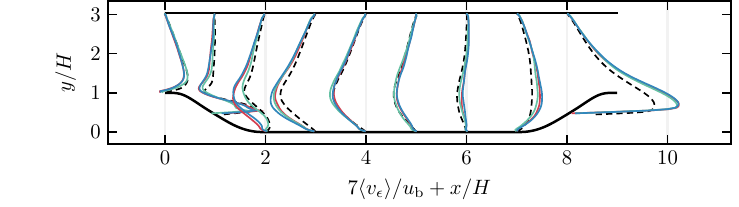}
    \caption{Mean streamwise (top) and wall-normal (bottom) volume-filtered velocity obtained with the VF-WMLES and by explicit volume-filtering of the DNS data of \citet{Krank2018} for the fine fluid mesh configuration. The VF-WMLES is carried out with the non-linear model (NL), the Vreman model (VRE), and the mixed model (Mix).}
    \label{fig:meanprofiles126}
\end{figure}

\begin{figure}
    \centering
    \includegraphics[scale=1.2]{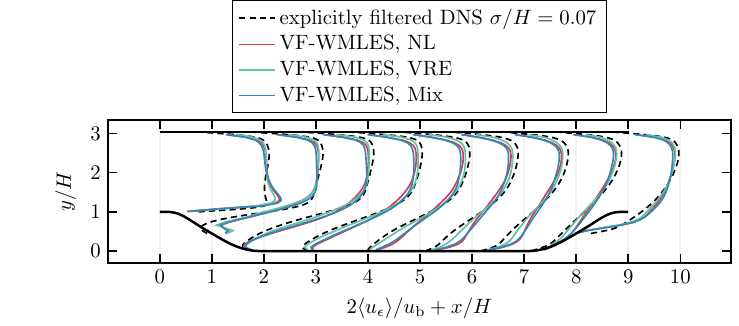}
    \includegraphics[scale=1.2]{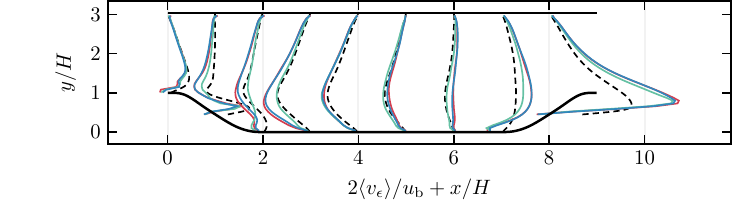}
    \caption{Mean streamwise (top) and wall-normal (bottom) volume-filtered velocity obtained with the VF-WMLES and by explicit volume-filtering of the DNS data of \citet{Krank2018} for the coarse fluid mesh configuration. The VF-WMLES is carried out with the non-linear model (NL), the Vreman model (VRE), and the mixed model (Mix).}
    \label{fig:meanprofiles64}
\end{figure}

\begin{figure}
    \centering
    \includegraphics[width=0.5\linewidth]{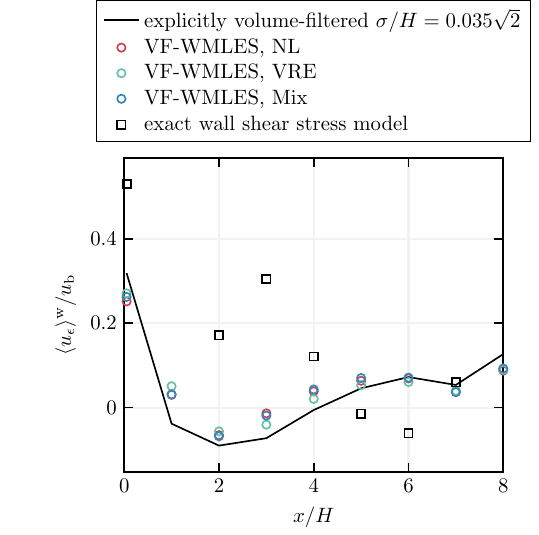}
    \caption{Comparison of the mean of the volume-filtered streamwise velocity that is enforced at the wall with the proposed VF-WMLES using the fine fluid mesh configuration and a classical wall model with the exact wall shear stress. }
    \label{fig:desiredvelocitywall126}
\end{figure}

\begin{figure}
    \centering
    \includegraphics[width=0.5\linewidth]{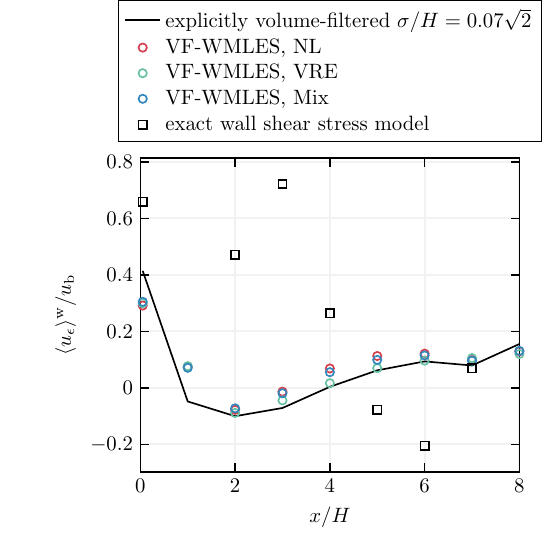}
    \caption{Comparison of the mean of the volume-filtered streamwise velocity that is enforced at the wall with the proposed VF-WMLES using the coarse fluid mesh configuration and a classical wall model with the exact wall-shear stress.}
    \label{fig:desiredvelocitywall64}
\end{figure}

\begin{table}[htb]
\caption{\label{tab:meshperiodichill} Fluid mesh information and filter width for the fine and coarse configuration of the periodic hill flow.}
\begin{ruledtabular}
\begin{tabular}{ccccc}
Name & $\overline{L}_y$ & $N_x\times N_y \times N_z$ & $\Delta x/H \times \Delta y/H \times \Delta z/H$ & $\sigma/H$\\
\hline
fine & $3.461H$ & $126 \times 98 \times 64$ & $0.0714 \times 0.0353 \times 0.0703$ & $0.035$ \\
coarse & $3.85H$ & $64 \times 55 \times 32$ & $0.141\times 0.07 \times 0.141$ & $0.07$ \\
\end{tabular}
\end{ruledtabular}
\end{table}
Figure \ref{fig:contourperiodichill} shows a snapshot of the absolute velocity of the VF-WMLES obtained with the fine fluid mesh. The numerical domain is larger than the physical domain but in the regions outside of the physical domain that have a sufficient distance from the wall, the magnitude of the volume-filtered velocity is very small. \\
As shown in section \ref{ssec:volumefilteredvelocityatwall}, approximating the volume filtered velocity at the wall with $u_{\epsilon,i}^{\mathrm{des}} = \sqrt{2}\sigma(\partial_nu_{\epsilon,i})^{\mathrm{w}}$ is accurate, even for the instantaneous velocity profiles of a turbulent channel flow. The velocity profiles of the periodic hill configuration, however, can have a negative streamwise velocity which may affect the accuracy of the model for the volume-filtered velocity at the wall. Therefore, the validity of the modeled desired volume-filtered velocity is tested. Figure \ref{fig:aprioriPeriodicHill} shows the mean streamwise velocity profiles as a function of the distance from the lower wall, $y-y_0$, where $y_0$ is the location of the lower wall, and the ratio of the volume-filtered velocity at the wall and the wall-normal gradient of the volume-filtered velocity at the wall for different $x$-positions and different filter widths. The DNS data is extracted from \citet{Krank2018}. It is observed that, except for the profile at $x/H=4.0$, the computed ratios lie within or close to the theoretical limits. At $x/H=4.0$, the mean streamwise velocity profile is close to zero up to a relatively large distance to the wall. Therefore, the volume-filtered gradient in the wall-normal direction is close to zero, which yields large deviations from the expected ratio. However, since the volume-filtered gradient in the wall-normal direction is very small, the resulting desired volume-filtered velocity at the wall is essentially zero with a small absolute error. These a priori results suggest that even in the complex separated flow over periodic hills, the proposed model for the volume-filtered velocity at the wall is relatively accurate. \\
Figure \ref{fig:meanprofiles126} shows the mean volume-filtered velocity in the streamwise and wall-normal direction obtained with the VF-WMLES and the fine fluid mesh configuration. The results are shown for different positions and compared to the explicitly volume-filtered DNS of \citet{Krank2018} with the same filter width. It should be noted, that since the mean velocity of the DNS is only known along lines in the wall-normal direction, the explicit volume-filtering is only carried out in the wall-normal direction. However, filtering in the wall-normal direction instead of all directions is expected to be negligible since the mean velocity varies much more rapid in the wall-normal direction than in the streamwise and spanwise direction. The profiles of the VF-WMLES are obtained with the non-linear model, the Vreman model, and the mixed model. It can be observed in figure \ref{fig:meanprofiles126} that the streamwise profiles obtained with the VF-WMLES accurately predict the explicitly volume-filtered DNS profiles in most of the regions. The most significant deviations of the streamwise profiles occur in the reattachment region near the lower wall. In this region the mean volume-filtered streamwise velocity is slightly overpredicted. The impact of the model for the subfilter stress tensor on the streamwise profiles is relatively minor but the best agreement is achieved with the Vreman model. \\
The wall-normal velocity is also predicted accurately with the exception at $x/H=8$, where the wall-normal velocity is oversestimated by the VF-WMLES with all of the three models for the subfilter stress tensor. At $x/H=2$, the VF-WMLES with the non-linear model and with the mixed model predict a different mean of the volume-filtered wall-normal velocity but with the Vreman model the explicitly volume-filtered DNS is accurately captured. Apart from $x/H=2$, the model for the subfilter stress tensor has a minor influence on the mean of the volume-filtered wall-normal velocity. \\
The VF-WMLES are also carried out with the coarse fluid mesh configuration with the twice the filter width than in the fine fluid mesh configuration. The resulting mean profiles of the streamwise and wall-normal volume-filtered velocity are shown in figure \ref{fig:meanprofiles64}. Although larger deviations from the explicitly volume-filtered DNS are observed than with the fine fluid mesh, good agreement of the profiles can be observed for most of the regions in the domain. The influence of the model for the subfilter stress tensor is stronger than with the smaller filter width and the best agreement is achieved with the Vreman model. 
The volume-filtering framework allows to formulate the correct slip and penetration boundary conditions dependent on the filter width, which can be obtained by explicitly volume-filtering the DNS velocity field at the wall. The explicitly volume-filtered velocity at the wall can be compared to the desired volume-filtered velocity enforced in the VF-WMLES to assess the accuracy of the wall model. The possibility of formulating such a reference solution does not exist in wall shear stress models, because applying a Neumann boundary condition on the filtered velocity based on the unfiltered wall shear stress is an ad hoc intervention that, by construction, predicts the correct bulk velocity but lacks any physical justification. \\
Figure \ref{fig:desiredvelocitywall126} shows the mean of the explicitly volume-filtered DNS velocity at the wall compared to the mean of the desired volume-filtered velocity of the VF-WMLES with the fine fluid mesh configuration for different positions in the periodic hill channel. Note that the factor $\sqrt{2}$ arises in the filter width because the desired volume-filtered velocity exists at the surface markers with a filter width $\sqrt{2}\sigma$ (see section \ref{ssec:modelingsi} for details). Figure \ref{fig:desiredvelocitywall126} shows the velocity at the wall that a wall shear stress model would predict if the filtered velocity and the wall shear stress are known exactly. Assuming that the first fluid mesh cell center lies in a distance $\sqrt{2}\sigma/2$ from the wall and the corresponding velocity in this cell is $u_{\epsilon,1}=u_{\epsilon}|_{y=y_0+\sqrt{2}\sigma/2}$, extrapolation with the wall shear stress results in the following prediction of the mean of the volume-filtered streamwise velocity at the wall
\begin{align}
    \langle u_{\epsilon} \rangle^{\mathrm{w}} = \langle u_{\epsilon,1}\rangle - \sqrt{2}\sigma \tau_{\mathrm{w}}/(2\mu_{\mathrm{f}}),
\end{align}
which we refer to as the exact wall shear stress model. \\
The largest slip velocity occurs at the crest of the hill, i.e., $x/H=0$. After the crest of the hill, a recirculation region follows with a negative slip velocity. After the recirculation region, a small positive slip velocity is observed for the rest of the channel. The desired volume-filtered velocity with the proposed VF-WMLES predicts the explicitly volume-filtered DNS velocity at the wall well along the whole length of the periodic hill configuration. The influence of the model for the subfilter stress tensor on the desired volume-filtered velocity is very minor. The exact wall shear stress model predicts large deviations from the explicitly volume-filtered DNS velocity and does not even show the correct trend along the channel. The reason for the poor agreement is that wall shear stress models enforce the boundary condition of the unfiltered velocity, i.e., the unfiltered wall-normal velocity gradient, on the filtered velocity. With the volume-filtering framework, the correct bulk velocity can be predicted even with a much smaller wall-normal velocity gradient because of the momentum source $s_i$ in the volume-filtered momentum equation \eqref{eq:reducedNSEmomentumwall}. Since $s_i$ is missing in wall-shear stress models, an artificially large wall shear stress must be imposed to recover the correct bulk velocity. \\
In figure \ref{fig:desiredvelocitywall64} the mean volume-filtered velocities at the wall are compared for the coarse fluid mesh configuration. The VF-WMLES approximates the mean of the explicitly volume-filtered DNS velocity at the wall well, which demonstrates that the VF-WMLES works accurately even at coarse resolutions of complex flow configurations with separation. The desired velocity is not very sensitive to the choice of the model for the subfilter stress tensor. The exact wall shear stress model provides poor predictions of the mean of the explicitly volume-filtered DNS velocity at the wall. \\
In studies using WMLES it is common to compare the friction and pressure coefficient predicted with the WMLES to the friction and pressure coefficient obtained from the corresponding DNS. However, the friction coefficient contains the unfiltered wall-normal velocity gradient and the pressure coefficient contains the unfiltered pressure, quantities that cannot be accessed in an LES, in which filtered quantities are determined. In the proposed VF-WMLES with the PC-IBM, the total hydrodynamical stress vector on the boundary, $\Sigma_i$, is computed at every surface marker (see section \ref{ssec:modelingsi} for details), which allows to compute the distribution of the total stress at the boundary and the resulting force.

\section{Conclusions}
\label{sec:conclusions}
In the present paper, the concept of wall-modeled large eddy simulations (WMLES) is revisited using the concept of volume-filtering. Volume-filtered quantities are well defined throughout the domain, even close to the boundaries of the flow domain, where classical LES filtering is not applicable. Recent advances in theory \cite{Hausmann2024a} and modeling \cite{Hausmann2024b} with the volume-filtering framework enable a new perspective on wall-modeling, which we refer to as volume-filtered WMLES (VF-WMLES). \\
We show that a non-zero slip and penetration velocity at the wall is a direct consequence of volume-filtering the flow. In contrast to wall shear stress models, the VF-WMLES framework allows to formulate a physically consistent reference solution for the volume-filtered velocity at the wall that is obtained by explicitly volume-filtering the DNS velocity field. With this, wall models can be assessed if they can predict this velocity at the wall. We propose a model to predict the volume-filtered velocity at the wall that is based on the ratio of the volume-filtered velocity and the wall-normal gradient of the volume-filtered velocity, which is assumed to be equal to the filter width. The model is shown to accurately predict the volume-filtered velocity at the wall in a priori tests based on existing DNS databases and in a posteriori tests with a turbulent channel flow and a turbulent flow over periodic hills.  \\
The desired volume-filtered velocity at the wall is enforced with the recently proposed PC-IBM, a modeling framework based on the volume-filtering concept, that enables a consistent representation of arbitrarily shaped boundaries on a relatively coarse Cartesian fluid mesh \cite{Hausmann2024b}. \\
The VF-WMLES is applied to a turbulent channel flow and a turbulent flow over periodic hills using different resolutions and different models for the subfilter stress tensor. The mean volume-filtered velocities are predicted with good accuracy, although there are variations between the results using different models for the subfilter stress tensor. Since the mean volume-filtered velocity at the wall is recovered well by the VF-WMLES, it seems that the major cause for the relatively small remaining deviations of the mean velocity is the model for the subfilter stress tensor. \\
With the VF-WMLES, we propose a framework with great potential to overcome common issues of existing wall-models such as their formulation based on mean quantities, the assumption of zero pressure gradient, or the necessity of a non-uniform filter that causes commutation errors. The possibility to formulate a ground truth of the volume-filtered velocity enables a priori and a posteriori testing of the wall boundary conditions for wall-models proposed in the future, which is not possible with the commonly used wall-shear stress models.

\begin{acknowledgments}
This research was funded by the Deutsche Forschungsgemeinschaft (DFG, German Research Foundation) \textemdash Project-ID 457509672. 
\end{acknowledgments}


%

\appendix
\section{Derivation of the limits of the velocity to velocity gradient ratio}
\label{ap:limitsVTGR}
Assuming that the wall is located at $x_{\mathrm{n}}=0$, where $x_{\mathrm{n}}$ is the wall normal coordinate, a volume-filtered element of the series expansion equation \eqref{eq:rootexpansion} using a one-dimensional Gaussian filter kernel, $g_{1\mathrm{D}}$, with a filter width, $\sigma$, is given as 
\begin{align}
   \int\displaylimits_0^{\infty}g_{1\mathrm{D}}(x_{\mathrm{n}})x_{\mathrm{n}}^{1/k} \mathrm{d}x_{\mathrm{n}} =\dfrac{2^{-1+1/(2k)}}{\sqrt{\pi}}\sigma^{1/k}\Gamma\left(\dfrac{1+k}{2k}\right),
\end{align}
where $\Gamma$ is the Gamma-function, i.e., a real number, and $k\in \mathbb{N}$. The volume-filtered derivative in $x_{\mathrm{n}}$-direction is given as
\begin{align}
    \int_{0}^{\infty}g_{1\mathrm{D}}(x_{\mathrm{n}}) \dfrac{1}{k}x_{\mathrm{n}}^{-1+1/k} \mathrm{d}x_{\mathrm{n}} = \dfrac{2^{-1/2+1/(2k)}}{\sqrt{\pi}}\sigma^{-1+1/k}\Gamma\left(1+\dfrac{1}{2k}\right).
\end{align}
The ratio of the volume-filtered profile and its volume-filtered derivative in $x_{\mathrm{n}}$-direction is given as
\begin{align}
    \dfrac{\int_{0}^{\infty}g_{1\mathrm{D}}(x_{\mathrm{n}}) x_{\mathrm{n}}^{1/k} \mathrm{d}x_{\mathrm{n}}}{\int_{0}^{\infty}g_{1\mathrm{D}}(x_{\mathrm{n}}) \dfrac{1}{k}x_{\mathrm{n}}^{-1+1/k} \mathrm{d}x_{\mathrm{n}}} = \dfrac{\Gamma\left(\dfrac{1+k}{2k}\right)}{\sqrt{2}\Gamma\left(1+\dfrac{1}{2k}\right)}\sigma.
\end{align}
The factor before $\sigma$ is a monotonously increasing function for $k\ge1$ with the limit
\begin{align}
    \lim_{k\rightarrow\infty}\left(\dfrac{\Gamma\left(\dfrac{1+k}{2k}\right)}{\sqrt{2}\Gamma\left(1+\dfrac{1}{2k}\right)} \right) = \sqrt{\dfrac{\pi}{2}}\approx 1.25331.
\end{align}
For $k=1$, the factor becomes
\begin{align}
    \dfrac{\Gamma\left(1\right)}{\sqrt{2}\Gamma\left(\dfrac{3}{2}\right)} = \sqrt{\dfrac{2}{\pi}}\approx 0.79788.
\end{align}
Therefore, if the velocity can be represented by the series expansion equation \eqref{eq:rootexpansion}, the following limits can be formulated for the ratio of the volume-filtered velocity to the wall-normal gradient of the volume-filtered velocity
\begin{align}
    \sqrt{\dfrac{2}{\pi}}\sigma\le \dfrac{u_{\epsilon,\alpha}^{\mathrm{w}}}{(\partial_nu_{\epsilon,\alpha})^{\mathrm{w}}} \le \sqrt{\dfrac{\pi}{2}}\sigma.
\end{align}

\end{document}